\documentclass[conference]{IEEEtran}
\IEEEoverridecommandlockouts
\usepackage{cite}
\usepackage{amsmath,amssymb,amsfonts}
\usepackage{textcomp}
\usepackage[a4paper, total={184mm,239mm}]{geometry}
\usepackage[utf8]{inputenc} %
\usepackage[T1]{fontenc}    %
\usepackage{pifont}
\usepackage[colorlinks,
    linkcolor=red,citecolor=citecolor,urlcolor=blue]{hyperref}       %

\usepackage{graphicx}
\usepackage[table,xcdraw]{xcolor}
\usepackage{threeparttable}
\usepackage{multirow}
\usepackage{amsmath}
\usepackage{bm}
\usepackage{float}
\usepackage{amsthm}
\usepackage{soul}
\usepackage[noend]{algpseudocode}
\usepackage{enumitem} %
\usepackage[subrefformat=parens,labelformat=parens]{subfig}

\usepackage{xspace}

\usepackage{multirow}

\definecolor{citecolor}{RGB}{34,139,34}
\definecolor{mydarkblue}{rgb}{0,0.08,1}
\definecolor{mydarkgreen}{rgb}{0.02,0.6,0.02}
\definecolor{mydarkred}{rgb}{0.8,0.02,0.02}
\definecolor{mydarkorange}{rgb}{0.40,0.2,0.02}
\definecolor{mypurple}{RGB}{111,0,255}
\definecolor{myred}{rgb}{1.0,0.0,0.0}
\definecolor{mygold}{rgb}{0.75,0.6,0.12}
\definecolor{myblue}{rgb}{0,0.2,0.8}
\definecolor{mydarkgray}{rgb}{0.,0.2,0.2}

\definecolor{lightred}{RGB}{255,235,235}
\definecolor{lightgreen}{RGB}{235,255,235}
\definecolor{lightblue}{RGB}{235,235,255}
\definecolor{lightcyan}{RGB}{235,255,255}
\definecolor{lightmagenta}{RGB}{255,235,255}
\definecolor{lightyellow}{RGB}{255,255,235}

\definecolor{qxkcolor}{RGB}{215,235,255}
\definecolor{softmaxcolor}{RGB}{230,235,255}
\definecolor{probxvcolor}{RGB}{255,255,235}

\definecolor{topkcolor}{RGB}{255,235,235}
\definecolor{zecolor}{RGB}{255,255,235}
\definecolor{dynacolor}{RGB}{235,255,255}

\definecolor{reviewcolor}{RGB}{0,0,200}

\DeclareMathOperator*{\argmax}{argmax}

\theoremstyle{plain}

\theoremstyle{definition}

\algdef{SE}[DOWHILE]{Do}{doWhile}{\algorithmicdo}[1]{\algorithmicwhile\ #1}%

\newcommand{\squishlist}{
 \begin{list}{$\bullet$}
  { \setlength{\itemsep}{0pt}
     \setlength{\parsep}{3pt}
     \setlength{\topsep}{3pt}
     \setlength{\partopsep}{0pt}
     \setlength{\leftmargin}{1.5em}
     \setlength{\labelwidth}{1em}
     \setlength{\labelsep}{0.5em} } }
     
\newcommand{\squishend}{
  \end{list}  }

\newcommand{\name}{\texttt{MAPS}\xspace}
\newcommand{\invdes}{\texttt{MAPS-InvDes}\xspace}
\newcommand{\data}{\texttt{MAPS-Data}\xspace}
\newcommand{\train}{\texttt{MAPS-Train}\xspace}

\def\BibTeX{{\rm B\kern-.05em{\sc i\kern-.025em b}\kern-.08em
    T\kern-.1667em\lower.7ex\hbox{E}\kern-.125emX}}

\begin{document}

\title{\texttt{MAPS}: \underline{M}ulti-Fidelity \underline{A}I-Augmented \underline{P}hotonic \underline{S}imulation and Inverse Design Infrastructure 
}

\author{
Pingchuan Ma$^1$,
Zhengqi Gao$^2$,
Meng Zhang$^3$,
Haoyu Yang$^4$,
Mark Ren$^4$,\\
Rena Huang$^3$,
Duane S. Boning$^2$,
Jiaqi Gu$^{1\dagger}$\\
$^1$Arizona State University, $^2$Massachusetts Institute of Technology, $^3$Rensselaer Polytechnic Institute, $^4$NVIDIA \\ {\small $\dagger$jiaqigu@asu.edu}
}

\maketitle

\begin{abstract}
Inverse design has emerged as a transformative approach for photonic device optimization, enabling the exploration of high-dimensional, non-intuitive design spaces to create ultra-compact, high-performance devices, advancing photonic integrated circuits (PICs) in computing and interconnects. 
However, practical challenges, such as suboptimal device performance compared to manual designs, limited manufacturability, high sensitivity to variations, computational inefficiency, and lack of interpretability, have hindered its adoption in commercial hardware.
Recent advancements in AI-assisted photonic simulation and design offer transformative potential, accelerating simulations and design generation by orders of magnitude over traditional numerical methods.
Despite these breakthroughs, the lack of an open-source, standardized infrastructure and evaluation benchmark limits accessibility and cross-disciplinary collaboration.
To address this, we introduce \texttt{MAPS}\footnote{We open-source our codes at \href{https://github.com/ScopeX-ASU/MAPS}{https://github.com/ScopeX-ASU/MAPS}}, a multi-fidelity AI-augmented photonic simulation and inverse design infrastructure designed to bridge this gap. 
\texttt{MAPS} features three synergistic components:
\ding{202}~\data: A dataset acquisition framework for generating multi-fidelity, richly labeled device designs using intelligent sampling strategies, providing high-quality data for AI-for-optics research.
\ding{203}~\train: A flexible AI-for-photonics training framework offering a hierarchical data loading pipeline, customizable model construction, support for data- and physics-driven losses, and comprehensive evaluation metrics.
\ding{204}~\invdes: An advanced adjoint method-based inverse design toolkit that abstracts complex physics but exposes flexible optimization steps, integrates pre-trained AI models, and incorporates fabrication-aware variation models for real-world applicability.
This infrastructure \texttt{MAPS} provides a unified, open-source platform for developing, benchmarking, and advancing AI-assisted photonic design workflows, accelerating innovation in photonic hardware optimization and scientific machine learning.
\end{abstract}

\section{Introduction}
The field of photonics has witnessed significant advancements in recent years, driven by the increasing demands for ultra-compact and high-performance devices in photonic integrated circuits (PICs). These devices are pivotal in addressing the growing needs of modern computing and optical interconnect technologies. Traditional design methodologies, which heavily rely on manual design heuristics and iterative optimization, often fail to fully exploit the complex, high-dimensional design spaces inherent in photonic systems. To overcome these limitations, inverse design has emerged as a powerful paradigm, enabling the systematic exploration of unconventional and non-intuitive device configurations.

Inverse design leverages computational optimization to generate photonic device designs that meet specific performance criteria. By navigating through vast design spaces, this approach has demonstrated remarkable potential in achieving designs that are both compact and highly efficient. However, despite its promise, the practical adoption of inverse design in real-world applications remains constrained by several challenges. These include suboptimal performance compared to manual designs, limited manufacturability due to non-ideal fabrication processes\cite{wang2019robust, hammond2021photonic, hughes2019forward}, high sensitivity to variations\cite{wang2011robust, SCHEVENELS20113613}, computational inefficiencies, and the lack of interpretability in design outcomes. Such barriers hinder the transition of inverse-designed devices from research laboratories to commercial hardware.

Recent advancements in artificial intelligence (AI) have opened new avenues for photonic simulation and design. 
AI-driven methods offer unprecedented acceleration in simulations and optimization, achieving speedups that are orders of magnitude faster than traditional numerical techniques\cite{NN_ICLR2021_Li, NN_ICLR2023_Tran, NN_MICCAI2015_Ronneberger, zhang2024sinenet, NP_NeurIPS2022_Gu, zhu2024Pace}. These advancements have significantly enhanced the feasibility of leveraging inverse design for practical photonic applications. Nonetheless, the broader adoption of AI-assisted approaches in photonics is impeded by the absence of an open-source, standardized infrastructure and evaluation benchmarks. This lack of standardized tools limits accessibility for researchers and prevents effective cross-disciplinary collaboration.

To address these critical gaps, we introduce \texttt{MAPS} (Multi-fidelity AI-augmented Photonic Simulation and inverse design), a comprehensive infrastructure that bridges the divide between theoretical advancements and practical implementation. MAPS is designed to be a unified, open-source platform that supports the entire workflow of AI-assisted photonic design. It integrates three core components:
\begin{itemize}[leftmargin=*]
\setlength{\itemindent}{0.5em}
\vspace{-3pt}
    \item \textbf{\data:} A dataset acquisition framework that allows flexible sampling strategies to generate multi-fidelity, richly labeled photonic device designs. 
    
    \item \textbf{\train} A flexible training framework specifically for AI-for-photonics research. This module supports hierarchical data pipelines, customizable model architectures, and physics-driven loss functions while offering comprehensive evaluation metrics.
    
    \item \textbf{\invdes} An advanced toolkit for adjoint inverse optimization that abstracts the complexities of photonic physics while incorporating AI models and fabrication-aware variation models for real-world applicability.
\end{itemize}

\section{Preliminary}
\subsection{Conventional Photonic Device Inverse Design}
\label{sec:conventional_adjoint_method}
The conventional adjoint-method-based photonic device inverse design can be formulated as follows:
\begin{equation}
\small
\label{eq:invdesFormulation}
\begin{gathered}
\theta^* = \argmax_{\theta \in \Theta}~F(\epsilon(\theta)|\lambda_c, J), \\
\text{s.t.}~~
\Bar{\rho} = (\mathcal{G}\circ \mathcal{P})(\theta)\\ 
\epsilon = \epsilon_v + (\epsilon_s-\epsilon_v)\cdot \Bar{\rho}\\
\mathcal{P}: \theta \in \mathbb{R}^N \to \rho \in \{0, 1\}^{N^x \times N^y}\\
\mathcal{G}: \rho \in \{0, 1\}^{N^x \times N^y} \to \Bar{\rho} \in [0, 1]^{N^x \times N^y} \\
\end{gathered}
\end{equation}
Here, $\lambda_c$ and $J$ represent the wavelength of interest and the injected light source, respectively. The design variables, $\theta$, first parameterize the design patterns through $\mathcal{P}$. 
This is followed by a combination of projections $\mathcal{G}$ applied to the parameterized pattern $\rho$ to generate the final device design, $\Bar{\rho}$, in which various heuristics have been proposed in prior work to minimize the gap between the numerically optimized device performance and the post-fabrication performance\cite{khoram2020controlling, schubert2022inverse, DID_ACSP2020_Chen, gershnabel2022reparameterization, ma2024boson1understandingenablingphysicallyrobust}.
The design patterns $\Bar{\rho}$ are then passed into a numerical solver (e.g., FDFD or FDTD) to perform forward and adjoint simulations. Finally, using the calculated adjoint gradient, backpropagation updates the design variables, enabling iterative optimization of the design.

\subsection{Machine Learning-Assisted Photonic Inverse Design}

Machine learning-assisted photonic inverse design methods can generally be categorized into two classes: \emph{generative} methods and \emph{predictive} methods, for now, \texttt{MAPS} only support predictive models while the generative models are left for future development.

The predictive method focuses on approximating the response of given design patterns. Traditionally, these responses are computed using numerical solvers that solve Maxwell's equations in either the time domain (FDTD) or the frequency domain (FDFD). Previous works \cite{zhu2021phase, Lin:22} have treated forward simulations as a black box, directly predicting specific quantities such as S-parameters. Other studies\cite{NP_NeurIPS2022_Gu, zhu2024Pace, ma2024pic2osimphysicsinspiredcausalityawaredynamic} have attempted to predict the electric or magnetic field distributions for a given design.

Despite the progress made, most efforts in predictive ML-assisted photonic inverse design to date have primarily concentrated on model development to achieve better approximations, among which the datasets, problem formulations, and evaluation metrics vary, resulting in difficulty in direct comparison.

\section{Proposed \texttt{MAPS} Infrastructure}
\begin{figure}
    \centering
    \includegraphics[width=0.95\columnwidth]{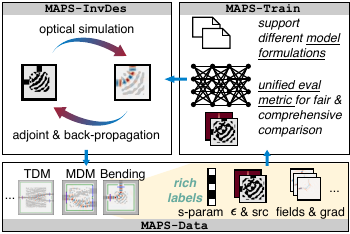}
    \caption{Illustration of \name infrastructure based on three synergistic sub-modules: \data, \invdes, and \train, targeting AI-assisted photonic simulation and inverse design.
    }
    \label{fig:totalFramework}
\end{figure}

Figure~\ref{fig:totalFramework} illustrates the main framework of the \name infrastructure, which comprises three key components. The first is a dataset, \data, featuring a balanced and more realistic data distribution. 
It covers a diverse range of photonic devices and includes rich labels to support universal training and evaluation of various AI-based PDE solvers. 
The second component is a PDE solver training framework, \train, which enables the training of AI models with different formulations and provides a unified, comprehensive evaluation of their performance. 
Finally, the inverse design framework, \invdes, offers high flexibility and user-friendliness, seamlessly integrating with AI-based PDE solvers to streamline the design process.
\subsection{\texttt{MAPS-Data}: A Dataset Balanced, More Reasonable Data Distribution and Rich Label Supporting Universal Training and Throughout Evaluation}
\label{sec:maps-data}
Data is critical to AI-assisted photonic device simulation and inverse design, e.g., performance prediction, forward field simulation, and inverse structure generation.
An ideal dataset for training models in photonic inverse design should possess the following properties: (1) it includes representative device structures that comprehensively cover the vast inverse design space, and (2) it should include detailed, informative labels to fully characterize the state of each sample and provide strong supervision to the model learning process. 
However, creating such an ideal dataset presents significant challenges. 
Given the high cost of data acquisition through numerical simulation and high-dimensional design space, the scarcity of large-scale, high-quality labeled data becomes a significant barrier to hindering model training and generalization. 
Efficient sampling strategies need to be explored to facilitate better model learning while maintaining high data acquisition efficiency; 
Simply increasing data volume is prohibitively costly and thus not practical in resolving the generalization dilemma for current AI models.

\begin{figure}
    \centering
    \includegraphics[width=0.75\columnwidth]{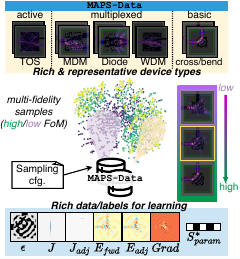}
    \caption{\data framework with various devices and sampling strategies. }
    \label{fig:dataFramework}
    \vspace{-10pt}
\end{figure}

To address this fundamental challenge, we propose \data, as shown in Fig.~\ref{fig:dataFramework}, a dataset acquisition framework that features: (1) a flexibly configurable scheme to support the exploration of diverse data collection strategies; 
(2) support the generation of rich, informative labels that thoroughly characterize each sample's state and figure-of-merits (FoMs) in the space that is interested specifically in inverse design's usage; 
(3) support of multi-fidelity simulation data generation that enables a trade-off between the data acquisition cost and dataset quality.
and (4) covers a wide variety of photonic devices, ranging from simple to complex, single-function to multiplexed, and passive to active components.
Next, we will elaborate on the unique features of \data and delve into their significance to AI-assisted photonic device design.

\subsubsection{Flexibly Configurable Data Sampling Strategy}
Given a limited data acquisition cost budget, it is critical to intelligently sample the most representative design points for simulation and label generation.
The actual device structures encountered by an adjoint gradient-based inverse device optimizer, as discussed in Section~\ref{sec:conventional_adjoint_method}, encompass soft structure patterns with low FoMs in the early stage and gradually transition to hard binarized structure with high FoMs when it converges.
However, most prior work usually predefines a design space with tunable design knots and randomly sample structures from the space,
This strategy only includes binarized design patterns~\cite{NP_NeurIPS2022_Gu, naseri2021generative} with low FoMs and often fails to train a generalizable model to support inverse design.

A natural question arises: \emph{How to strategically select data samples to approach the real distribution that the AI model encountered in the actual query during device inverse design?} 
To facilitate the exploration of alternative sampling strategies, \data provides a flexibly configurable framework that allows users to easily customize sampling approaches during data acquisition. 
\invdes model is seamlessly integrated with \data to enable an optimization-aware data sampling approach.
When integrated with \train, the sampled dataset can be used to train various AI models to evaluate its effectiveness using multiple device-specific metrics, enabling fair and comprehensive comparisons between different strategies.

\subsubsection{Rich Labels for Multi-task Learning and Evaluation}~
The rapid development of diverse machine learning methods for addressing the forward simulation and inverse design of photonic devices has introduced significant variability in their label requirements. 
For instance, black-box models often predict the S-parameters of specific devices, whereas models pursuing interpretability or adjoint gradients may predict electric or magnetic field distributions. 
To accommodate this diversity and fully leverage the information in a limited number of data samples, \data extracts rich labels from simulation results, including transmission, reflection, radiation, electrical/magnetic fields, adjoint gradient under a certain objective, Maxwell equation matrices, etc.
With rich labels for each data sample, we can support multi-task supervised and self-supervised learning, ensuring broad applicability and flexibility. 

\subsubsection{Varying-Fidelity Data for Robust, Efficient Multi-Fidelity Model Training}~

Multi-fidelity datasets provide device simulation results at varying quality levels, balancing the trade-off between data acquisition efficiency and quality. 
High-fidelity data, such as simulations of high-performance devices computed with fine mesh granularity, require significantly more computational resources, sometimes even require a time-consuming optimization process to find the structure, but offer more reliable supervision for training AI models.
In contrast, low-fidelity data with random patterns simulated with coarser mesh granularity are computationally inexpensive but less accurate and informative.
In numerical simulation literature, methods like Richardson extrapolation demonstrate how low-fidelity solutions can be systematically refined to approximate high-resolution results. 
This creates opportunities for AI models to integrate abundant low-fidelity data with a limited amount of high-fidelity data, enhancing generalization while maintaining data efficiency. 
By strategically leveraging these data sources, AI models can achieve robust performance with reduced data collection cost.
For generative AI models used in device inverse design, training datasets primarily consist of low-performance devices. High-performance devices, being the optimization target, are generally unavailable during training. 
This necessitates the use of low-FoM devices in the training phase, which further underscores the importance of the multi-fidelity property of \data.
To advance research in multi-fidelity modeling, \data provides \emph{paired device simulation data across fidelity levels}, enabling comprehensive exploration of fidelity-driven trade-offs in AI model training.

\subsubsection{A Broad Design Space of Representative Device Types}~
As illustrated in Fig.~\ref{fig:dataFramework}, \data encompasses a diverse range of inverse-designed photonic devices, from relatively simple structures like waveguide bends and crossings to more intricate designs, including optical diodes, wavelength division multiplexers (WDM), mode division multiplexers (MDM), and active thermo-optic switches (TOS). 
This extensive collection spans both single-function and multiplexing devices, as well as passive and active components, offering a comprehensive dataset with varying levels of complexity.
These devices represent diverse parameter dimensions in the device structure and operation, including variations in permittivity distribution, input light source wavelength, optical mode, and incident port locations. 
The dataset incorporates multiple design targets and performance metrics, making it a robust benchmarking platform for multi-task learning across a spectrum of challenges. 
By capturing devices of varying difficulty levels, \data supports the development of models capable of addressing complex, real-world design problems.

\subsection{\train: A Training Infrastructure Tailored for ML-assisted Photonic Device Inverse Design}
\label{sec:maps_train}
\begin{figure}
    \centering
    \includegraphics[width=0.98\columnwidth]{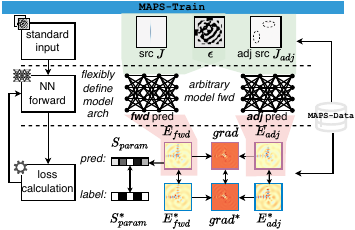}
    \caption{\train framework with standardized input, flexible neural network model definition, and diverse data-driven/physics-driven loss function calculation.}
    \label{fig:trainFramework}
\end{figure}

The success of AI-assisted photonic device simulation and inverse design hinges not only on the quality of the data but also on the model architecture and training procedures.
Previous work\cite{NN_IJDS2024_Li, NP_NeurIPS2022_Gu} has shown that improved model architectures and training methods, e.g., with physics-inspired modules, data augmentation, and loss functions, can significantly enhance optical simulation performance.
As AI models adopt diverse approaches to photonic device simulation and inverse design—particularly with advancements in physics-informed AI and multi-task learning—it becomes essential for training procedures to adapt accordingly. 
This includes supporting flexible training logic and enabling the construction of both data-driven and physics-driven loss functions.
A critical component of effective training is standardized model testing, which ensures consistent evaluation and facilitates fair comparisons across methods. 
Standardized metrics play a pivotal role in benchmarking models reliably. 
To meet these needs, an ideal training framework for photonic device design must satisfy the following criteria:
(1) Researchers should be able to define any model architecture conveniently, including ensembles or cascades of models, such as Tandem neural networks and generative adversarial networks (GANs).
(2) The framework should support flexible training workflows tailored to specific problem formulations, such as multi-task learning, distillation, pretraining and fine-tuning, multi-stage learning, and autoregression. 
Additionally, it must enable the incorporation of diverse loss functions, including data-driven (e.g., normalized MSE) and physics-driven (e.g., Maxwell equation residual) objectives.
(3) Evaluation should adhere to standardized metrics to ensure fair and consistent benchmarking. This includes general regression metrics and domain-specific criteria, such as S-parameter prediction error and adjoint gradient alignment.

To satisfy the above demand in model training, we develop \train, as illustrated in Fig.~\ref{fig:trainFramework}.
\train offers the following key features:
(1) \textbf{Hierarchical Data Loader Pipeline}: This pipeline partition data samples at the device level to prevent test set leakage and aggregates multiple sources or ports for each device, enabling effective data augmentation (e.g., superposition-based Mixup~\cite{NP_NeurIPS2022_Gu}).
(2) \textbf{Flexible Model Architecture Definition}: A flexible interface for defining model architectures, including different data encoding operators, model backbones, task-specific heads, and multi-model setups (e.g., Tandem neural networks) for both forward prediction and inverse generation.
(3) \textbf{Customizable Training Procedures}: The framework supports a wide range of training workflows, integrating data-driven loss functions (e.g., normalized MSE) and physics-driven loss components (e.g., Maxwell equation residual).
(4) \textbf{Comprehensive Model Evaluation}: Standardized evaluation metrics are included, ranging from regression errors to specialized metrics for inverse design, such as S-parameter prediction errors and adjoint gradient alignment.

\subsubsection{Flexibility and Extensibility}
Previous studies\cite{ma2024pic2osimphysicsinspiredcausalityawaredynamic, NP_NeurIPS2022_Gu, hughes2019wave},  have shown that physics priors and the manner in which a model is conditioned on its input can significantly impact the performance of AI solvers. \train provides a flexible framework for modifying network architectures, enabling the incorporation of additional ansatz into model designs to develop more advanced and effective AI solvers.

\subsubsection{A Unified Framework and Standardized Evaluation Metrics}
\label{sec:unifiedEvaluation}
Comparing AI solvers has been difficult due to variations in datasets, problem formulations, and evaluation metrics. \data addresses dataset inconsistencies, but differences in source representation and metrics—e.g., NMSE of S-parameters vs. normalized $L_2$ norm—complicate fair comparisons. 
\train resolves these issues by standardizing model inputs (permittivity $\epsilon$ and source $J$, as shown in Fig.~\ref{fig:totalFramework} and Fig.~\ref{fig:trainFramework}) and providing a comprehensive label set. It introduces gradient similarity as a key metric, critical for ensuring accurate optimization convergence. 
By integrating \data's unified datasets with \train's standardized metrics, this framework fosters fair benchmarking and collaboration in photonics machine learning.

\subsection{\texttt{MAPS-InvDes}: AI-Assisted Fabrication-Aware Adjoint Inverse Design Framework}  

Adjoint-based inverse design explores high-dimensional design spaces, enabling compact, high-performance photonic devices. However, its practical adoption is limited due to challenges such as:  
(1) Modest performance improvements in key metrics (e.g., insertion loss, bandwidth) compared to manual designs;  
(2) High computational demands from iterative simulations;  
(3) Performance degradation post-fabrication due to sensitivity to manufacturing variations;  
(4) Non-intuitive structures that hinder interpretability and trust.  

To address these challenges, advanced techniques are required for optimization, initialization, fabrication constraints, and accessible physics abstraction.  

\begin{figure}
    \centering
    \includegraphics[width=0.95\columnwidth]{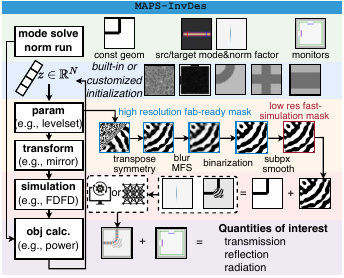}
    \caption{\invdes framework with full support of simulation and adjoint-based inverse design.}
    \label{fig:invdesFramework}
    \vspace{-10pt}
\end{figure}

\texttt{MAPS-InvDes}, illustrated in Fig.~\ref{fig:invdesFramework}, provides:  
(1) Easy definition of complex device structures;  
(2) Advanced topology parametrization for shape and size optimization;  
(3) Built-in options for initialization, constraints, and variation modeling;  
(4) Predefined and customizable optimization objectives with minimal physics expertise;  
(5) Integration with \train’s pre-trained AI solvers for efficient, AI-driven design.  

\subsubsection{Initialization}
Inverse design is inherently sensitive to initialization due to its high-dimensional, non-convex, discrete optimization landscape~\cite{ma2024boson1understandingenablingphysicallyrobust}. 
Poor initial conditions can lead to suboptimal convergence. 
To address this, \invdes provides predefined initializations designed for smooth convergence, enhancing the likelihood of reaching high-performance solutions. 
Additionally, it supports custom initialization strategies, allowing users to incorporate refined manual designs or domain-specific heuristics, e.g., encourage light transmission, to accelerate convergence and improve solution quality.

\subsubsection{Constraints and Reparametrization}
Manufacturability remains a significant bottleneck in translating optimized photonic designs into practical devices. 
\invdes mitigates this challenge by embedding critical fabrication-aware constraints, such as symmetry and minimum feature size (MFS) control through extensible, differentiable transformations. 
These constraints ensure that optimized designs adhere to real-world fabrication capabilities. 
Moreover, lithography and etching models are seamlessly integrated into the optimization loop, enabling end-to-end design processes that balance theoretical performance with practical manufacturability.

\subsubsection{Variation-Aware Inverse Design}
Robustness to manufacturing and operational variations is essential for real-world deployment of inverse-designed photonic devices. 
\invdes incorporates a differentiable lithography model~\cite{yang2024gpu} to ensure designs are not only optimized for ideal conditions but also resilient to fabrication-induced variations, e.g., over-etching, defocusing of lithography, and operation-time variations, e.g., laser wavelength shift and temperature drift. 
This variation-aware framework enables robust optimization within manufacturable subspaces. 
Additionally, it supports the integration of advanced AI-based fabrication models, allowing for adaptability to new manufacturing processes and materials, thereby extending the versatility of the design framework.

\subsubsection{Optimization Objectives}
Defining clear optimization objectives and their trade-offs is often a complex task in photonic inverse design, typically requiring specialized expertise in physics, such as far-field projection or eigenmode coefficient extraction. 
\invdes simplifies this process by offering a suite of predefined, composable optimization objectives, such as maximizing transmission efficiency or controlling far-field intensity distributions. 
This user-friendly framework allows researchers to flexibly specify design targets and approaches to balance multiple objectives, penalties, and regularization terms, enabling more accessible multi-objective optimization workflows.

\section{Case Studies}
\subsection{Results on Different Data Sampling Methods}
\begin{figure}
    \centering
    \vspace{-15pt}
    \subfloat[]{\includegraphics[width=0.47\columnwidth]{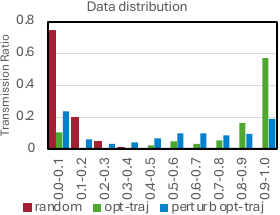}
    \label{fig:histogram}
    }
    \hspace{3pt}
    \subfloat[]{\includegraphics[width=0.47\columnwidth]{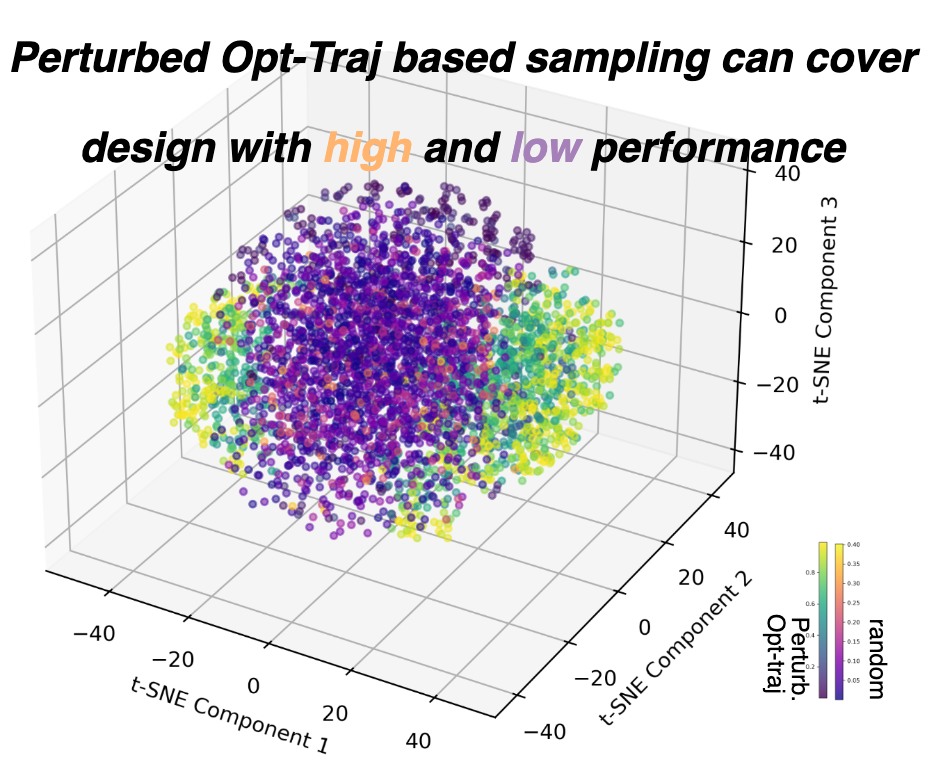}
    \label{fig:tSNE}
    }
    \caption{Comparison of sampling strategies: (a) Transmission ratio histogram for different strategies; (b) t-SNE showing separate distributions of low- and high-performance patterns, with perturbed opt-traj sampling covering both.}
    \label{fig:dataDistribution}
\end{figure}

\begin{figure}
    \centering
    \subfloat[]{\includegraphics[width=0.46\columnwidth]{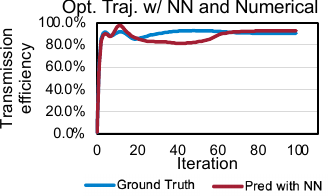}
    \label{fig:opt_traj_cmp}
    }
    \hspace{10pt}
    \subfloat[]{\includegraphics[width=0.41\columnwidth]{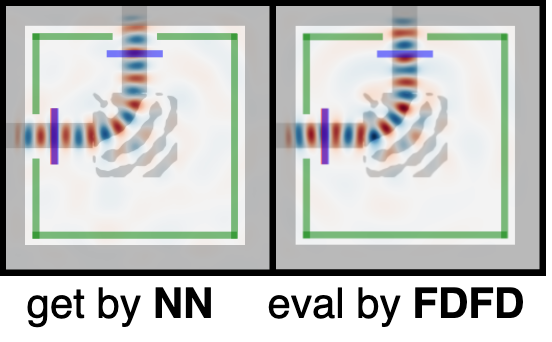}
    \label{fig:opt_traj_viz}
    }
    \caption{(a) Optimization trajectory driven by NN-predicted gradients. Transmission efficiency calculated based on NN-predicted fields and FDFD-simulated fields are shown for comparison.
    (b) The electrical field of the final design predicted by NN and verified by FDFD.
    }
    \label{fig:opt_traj_cmp_viz}
    \vspace{-5pt}
\end{figure}

In Table~\ref{tab:differentDataset}, the first case study using \name explores data sampling strategies versus random sampling. An ideal dataset must span the full design space, including low- and high-performance device patterns. 
Random sampling mainly yields low-performing devices, limiting the model's ability to predict high-performance patterns. Sampling along optimization trajectories captures diverse designs while avoiding redundancy in late-stage patterns.

Figure~\ref{fig:histogram} shows that random sampling of a bending waveguide produces low transmission efficiency (<10\%) in most samples. 
Optimization trajectory-based sampling includes a broader range but results in an unbalanced distribution. Perturbations during data selection address this imbalance, creating a more balanced distribution.
\begin{table}[]
\centering
\caption{Compared to a randomly sampled dataset, a model trained on a perturbed Opt-Traj dataset can provide better gradient similarity and more accurate field prediction.
}
\resizebox{1\columnwidth}{!}{%
\begin{tabular}{|c|c|c|c|c|}
\hline
models                 & dataset                                  & Train N-L2norm                 & Test N-L2norm                  & Grad Similarity                 \\ \hline
                       & \cellcolor[HTML]{EFEFEF}Perturb Opt-Traj & \cellcolor[HTML]{EFEFEF}0.1018 & \cellcolor[HTML]{EFEFEF}0.1881 & \cellcolor[HTML]{EFEFEF}0.42704 \\ \cline{2-5} 
\multirow{-2}{*}{FNO~\cite{NN_ICLR2021_Li}}  & random                                   & 0.1122                         & 0.7910                         & 0.08313                         \\ \hline
                       & \cellcolor[HTML]{EFEFEF}Perturb Opt-Traj & \cellcolor[HTML]{EFEFEF}0.4120 & \cellcolor[HTML]{EFEFEF}0.3401 & \cellcolor[HTML]{EFEFEF}0.27068 \\ \cline{2-5} 
\multirow{-2}{*}{UNet~\cite{NN_MICCAI2015_Ronneberger}} & random                                   & 0.5881                         & 0.8290                         & 0.02893                         \\ \hline
\end{tabular}
}
\label{tab:differentDataset}
\end{table}

To assess dataset strategies, we train two baselines on different datasets and evaluated them using normalized $L_2$ norm and gradient similarity. 
Results in Table~\ref{tab:differentDataset} show models trained on perturbed trajectory sampling outperform random sampling, achieving better prediction accuracy and gradient similarity.

\subsection{Gradient Computation Methods: Adjoint Fields vs. Auto-Diff}
\begin{table}[]
\centering
\caption{Comparison between different gradient calculation methods. The gradient calculated from the predicted forward and adjoint field is the most accurate.
}
\resizebox{0.7\columnwidth}{!}{%
\begin{tabular}{|c|c|c|}
\hline
models                 & Grad Method                              & Grad Similarity                \\ \hline
                       & AD-Black Box                             & 0.0511                         \\ \cline{2-3} 
                       & AD-Pred Field                            & 0.0552                         \\ \cline{2-3} 
\multirow{-3}{*}{FNO~\cite{NN_ICLR2021_Li}}  & \cellcolor[HTML]{EFEFEF}Fwd \& Adj Field & \cellcolor[HTML]{EFEFEF}0.4270 \\ \hline
                       & AD-Black Box                             & 0.0243                         \\ \cline{2-3} 
                       & AD-Pred Field                            & 0.0406                         \\ \cline{2-3} 
\multirow{-3}{*}{UNet~\cite{NN_MICCAI2015_Ronneberger}} & \cellcolor[HTML]{EFEFEF}Fwd \& Adj Field & \cellcolor[HTML]{EFEFEF}0.2707 \\ \hline
\end{tabular}
}
\label{tab:grad_method}
\end{table}

Predictive models in inverse design face a key challenge: gradient computation. 
There are two main approaches:
(1)~Black-box models or forward field predictors rely on auto-differentiation.
(2)~Field predictors compute gradients directly from forward and adjoint fields.
Table~\ref{tab:grad_method} highlights that using forward and adjoint fields yields higher gradient accuracy compared to auto-diff.
\subsection{Main Results}
\begin{table}[]
\centering
\caption{Comparison across different predictive baselines on different benchmark devices, the result format is Train N-L2norm/Test N-L2norm/Test gradient similarity.
}
\resizebox{\columnwidth}{!}{%
\begin{tabular}{|c|c|c|c|}
\hline
baselines  & bending        & crossing         & optical diode  \\ \hline
FNO~\cite{NN_ICLR2021_Li}        & 0.10/0.19/0.43 & 0.08/0.08/0.83   & 0.16/0.83/0.08 \\ \hline
F-FNO~\cite{NN_ICLR2023_Tran}      & 0.13/0.14/0.58 & 0.11/0.08/0.86   & 0.16/0.72/0.12 \\ \hline
UNet~\cite{NN_MICCAI2015_Ronneberger}       & 0.41/0.34/0.25 & 0.38/0.30/0.65    & 0.53/0.87/0.03 \\ \hline
NeurOLight~\cite{NP_NeurIPS2022_Gu} & 0.11/0.14/0.55 & 0.10/0.08/0.84 & 0.14/0.71/0.14 \\ \hline
           & MDM            & WDM              & TOS            \\ \hline
FNO~\cite{NN_ICLR2021_Li}        & 0.25/0.58/0.20 & 0.56/0.87/0.03   & 0.45/1.01/0.02 \\ \hline
F-FNO~\cite{NN_ICLR2023_Tran}      & 0.30/0.47/0.31 & 0.60/0.75/0.06   & 0.52/0.99/0.03 \\ \hline
UNet~\cite{NN_MICCAI2015_Ronneberger}       & 0.71/0.76/0.13 & 0.85/0.88/0.00   & 0.82/0.99/0.00    \\ \hline
NeurOLight~\cite{NP_NeurIPS2022_Gu} & 0.27/0.45/0.31 & 0.71/0.73/0.10    & 0.70/0.94/0.03  \\ \hline
\end{tabular}
}
\label{tab:mainResult}
\end{table}

We evaluate four baseline models—FNO, Factorized-FNO, UNet, and NeurOLight—on a set of benchmark photonic devices (Table~\ref{tab:mainResult}). 
The models are trained using the normalized $L_2$ norm between the predicted and ground truth electromagnetic fields ($E_z$, $H_x$, $H_y$), with magnetic fields $H_x$ and $H_y$ derived directly from the predicted $E_z$ distribution.

Among the evaluated models, NeurOLight consistently outperforms the others in both field prediction accuracy and gradient alignment, highlighting its effectiveness in capturing the underlying physical phenomena governing photonic device behavior. Its superior performance underscores the benefits of integrating physics knowledge in the model design and training.

However, despite these improvements, all models exhibit notable performance degradation when applied to more intricate device structures. 
This limitation suggests that current methods struggle to generalize across highly complex design spaces, pointing to the need for more advanced approaches. 
Potential future directions include incorporating advanced physics-informed architectures, leveraging multi-fidelity data integration, self-supervised learning, etc., to enhance the robustness and generalization capabilities of AI-assisted simulation.

\subsection{Integrate Neural Solver to \invdes}
As the last case study, we show the integration of the neural network model trained from \train to replace the numerical solver in photonic device inverse design, an ultimate interoperable demonstration of our \name infrastructure.
Figure~\ref{fig:opt_traj_cmp} shows the optimization trajectory purely driven by adjoint gradients calculated from NN-predicted simulation results and the device transmission evaluation at each iteration using FDFD as the ground truth.
The NN shows superior generalization at different points along the optimization trajectory.
The NN-predicted gradient aligns well with the true gradients, such that the optimized devices converge to a high-transmission structure, verified by the FDFD simulation.
This case study validates the feasibility of using AI-augmeted PDE solver to facilitate inverse design with promising solution quality and significant speedup.

\section{Conclusion}
In this work, we presented \name, a multi-fidelity, AI-augmented infrastructure for photonic simulation and inverse design, aimed at overcoming key challenges in data scarcity, computational efficiency, manufacturability, and design interpretability. 
By integrating \data, \train, and \invdes, this unified, open-source playground streamlines AI-assisted photonic design, enabling more efficient and standardized development, evaluation, and benchmarking.
Our case studies highlight \name’s ability to improve data diversity, model generalization, and optimization accuracy. 
\data generates balanced, multi-fidelity datasets that better reflect real-world device distributions, enhancing model training generalization. 
\train supports flexible model architectures and standardized evaluation metrics, ensuring fair comparisons and reproducibility. 
\invdes integrates fabrication-aware constraints and AI-driven solvers to streamline adjoint-based optimization and improve manufacturability and design efficiency.
Looking ahead, \name offers a solid foundation for advancing AI-driven photonic design, with potential extensions into generative models and more complex device architectures. 
By providing accessible, standardized tools, \name accelerates innovation in photonic hardware systems, paving the way for next-generation photonic integrated circuits in high-performance computing and optical communications.


\end{document}